\documentclass[12pt]{article}

\usepackage{epsf}
\usepackage[dvips]{graphicx}

\begin{document}

\begin{center}
{\bfseries FAST GENERATORS OF DIRECT PHOTONS}

\vskip 5mm

S.M. Kiselev$^{\dag}$\footnote{Talk at Baldin ISHEPP XIX, Dubna, September 29 - October 4 2008}

\vskip 5mm

{\small
{\it
Institute for Theoretical and Experimental Physics, Moscow, Russia
}
\\
$\dag$ {\it
E-mail: Sergey.Kiselev@cern.ch
}}
\end{center}

\vskip 5mm

\begin{center}
\begin{minipage}{150mm}
\centerline{\bf Abstract}
Three fast generators of direct photons in the central rapidity region of high-energy heavy-ion collisions have been presented 
The generator of prompt photons is based on a tabulation of $p+p(\bar p)$ data and binary scaling.
Two generators of thermal direct photons, for hot hadron gas (HHG) and quark-gluon plasma (QGP) scenarios,
assume the 1+1 Bjorken hydrodynamics. SPS and RHIC data can be fitted better by scenario with QGP. Predictions for the 
LHC energy have been made. The generators have been realized as macros for the ROOT analysis package.
\end{minipage}
\end{center}

\vskip 10mm

\section{Introduction}
Direct photons are photons not from particle decays. On the quark-gluon level three 
subprocesses dominate: Compton scattering $g q \rightarrow \gamma q$, 
annihilation $q \bar q \rightarrow \gamma g$ and bremsstrahlung emission
$q q(g) \rightarrow q q(g) \gamma$. Photons from initial hard NN collisions
are named prompt photons and are the main source at large $p_{t}$. 
They can be described by perturbative QCD (pQCD).
In the case of other limit - thermalized system of quarks and gluons,
the quark-gluon plasma (QGP), these photons are named thermal photons from QGP.

On the hadron level there are a lot of meson scattering channels:
$\pi \pi \rightarrow \rho \gamma, \pi \rho \rightarrow \pi \gamma,
\pi K \rightarrow K^* \gamma, K \rho \rightarrow K \gamma$, ... .
First two channels give most contribution. If hadron system is thermalized
these photons are named thermal photons from the hot hadron gas (HHG). 

Besides the meson rescatterings, photons from decays of short-living resonances, 
$\omega \rightarrow \pi \gamma$, $\rho \rightarrow \pi \pi \gamma$, $a_{1} \rightarrow \pi \gamma$, 
$\Delta \rightarrow N \gamma$..., make a contribution to direct photons. In the case when the
life time of a resonance is less than characteristic time of the nucleus-nucleus
collision it is difficult to reconstruct the resonance because the
decay hadron (e.g. $\pi$) can reinteract with surrounding medium especially
if this medium is dense.

Below three fast generators of direct photons are proposed: a generator of prompt
photons and two generators of thermal direct photons.

\section{Generator of prompt photons}

In the paper~\cite{PHENIX2006} all existing $p+p(\bar p)$ data on prompt photons in central
rapidity region have been presented as the function $(\sqrt{s})^{5}Ed^{3}\sigma^{pp}/d^{3}p=F(x_{T}), x_{T}=2p_{t}/\sqrt{s}$.
Using a tabulation of this dimensionless function one can estimate the prompt photon spectrum in nucleus-nucleus, 
A+B, collisions at the impact parameter $b$: $Ed^{3}N^{AB}(b)/d^{3}p = Ed^{3}\sigma^{pp}/d^{3}p \cdot AB \cdot T_{AB}(b)$,
where the nuclear overlapping function is defined as $T_{AB}(b)=N_{coll}(b)/\sigma^{pp}_{in}$,
where $N_{coll}(b)$ is the average number of binary NN collisions.
Nuclear effects (Cronin, shadowing) are ignored in this approach.
 
\begin{figure}[p]
\centerline{
\includegraphics[width=100mm,height=100mm]{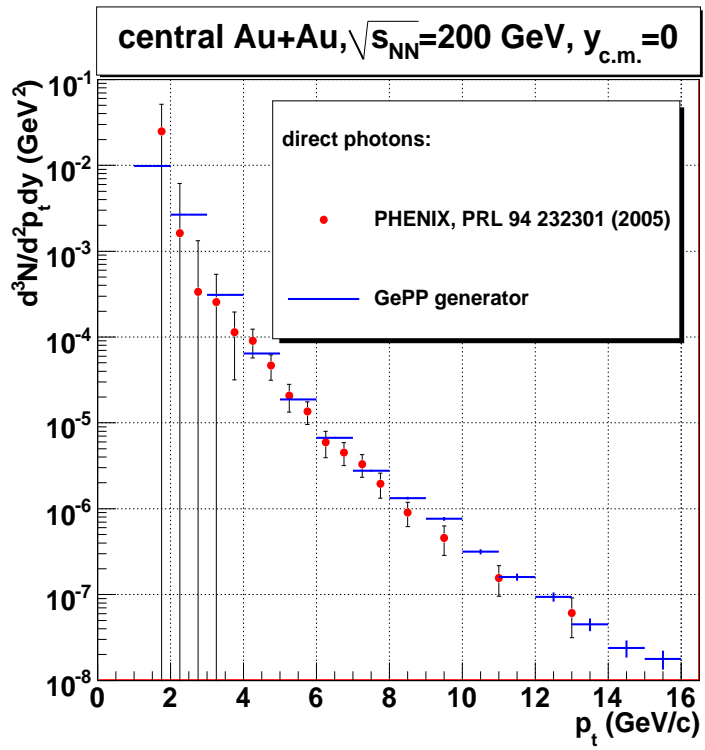}}
\caption{Direct photon spectrum in central Au+Au collisions at RHIC}
\label{fig:GePP-RHIC}
\end{figure}

Fig.~\ref{fig:GePP-RHIC} demonstrates reasonable agreement of existing RHIC data~\cite{PHENIX2005}
with results of the generator GePP.C realized as a macro for the ROOT analysis package ~\cite{ROOT}.
\begin{figure}[p]
\centerline{
\includegraphics[width=100mm,height=100mm]{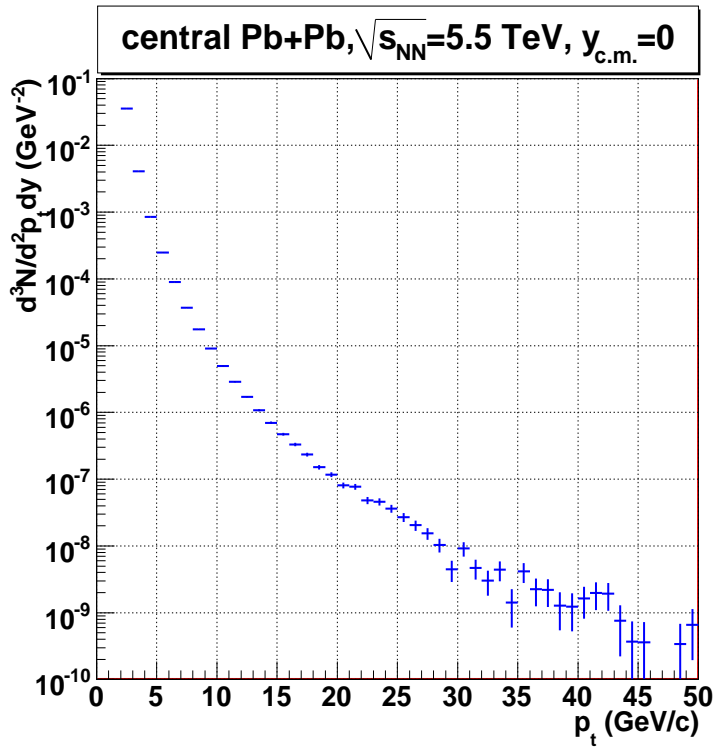}}
\caption{Prompt photon spectrum in central Pb+Pb collisions at LHC}
\label{fig:GePP-LHC}
\end{figure}
Predictions for prompt photons at LHC energy is presented in Fig.~\ref{fig:GePP-LHC}.

\section{Generators of thermal direct photons}
For fast generators the Bjorken hydrodynamics(BHD)~\cite{BHD} has been used.
It is assumed that during the ion-ion collision the system is mainly expanding in beam
direction in a boost-invariant way. Natural variables are the proper time $\tau=\sqrt{t^{2}-z^{2}}$ and rapidity. 
Thermodynamical variables (pressure, temperature, ...) do not depend on the rapidity but are functions of $\tau$. 
Viscosity and conductivity effects are neglected.

Main parameters are the initial thermalization time $\tau_{0}$ and temperature $T_{0}$. Photon yield is proportional to $\sim \tau_{0}^{2}$, the spectrum slope is controlled by $T_{0}$.
There is third parameter, temperature at freeze-out, $T_{f}$. Results depend weekly on it value. $T_{f}$=100 MeV has been used.

For this simple space-time evolution one can evaluate expression for photon spectrum with the photon emission rate as input
which are different for HHG and QGP scenarios. All needed formulas can be found in~\cite{PhysRep2002}.

\subsection{Generator for the HHG scenario}
The photon emission rates for processes $\pi \pi \rightarrow \rho \gamma, \pi \rho \rightarrow \pi \gamma$ and $\rho \rightarrow \pi \pi \gamma$ 
in the HHG scenario have been obtained in~\cite{Song} using the effective chiral Lagrangian theory with $\pi, \rho$ and $a_{1}$ mesons.
Later they were parametrized by formulas in~\cite{SongFai}. Rates for the channel $\omega \rightarrow \pi \gamma$ have been taken from~\cite{Kapusta}.

\begin{figure}[p]
\centerline{
\includegraphics[width=100mm,height=100mm]{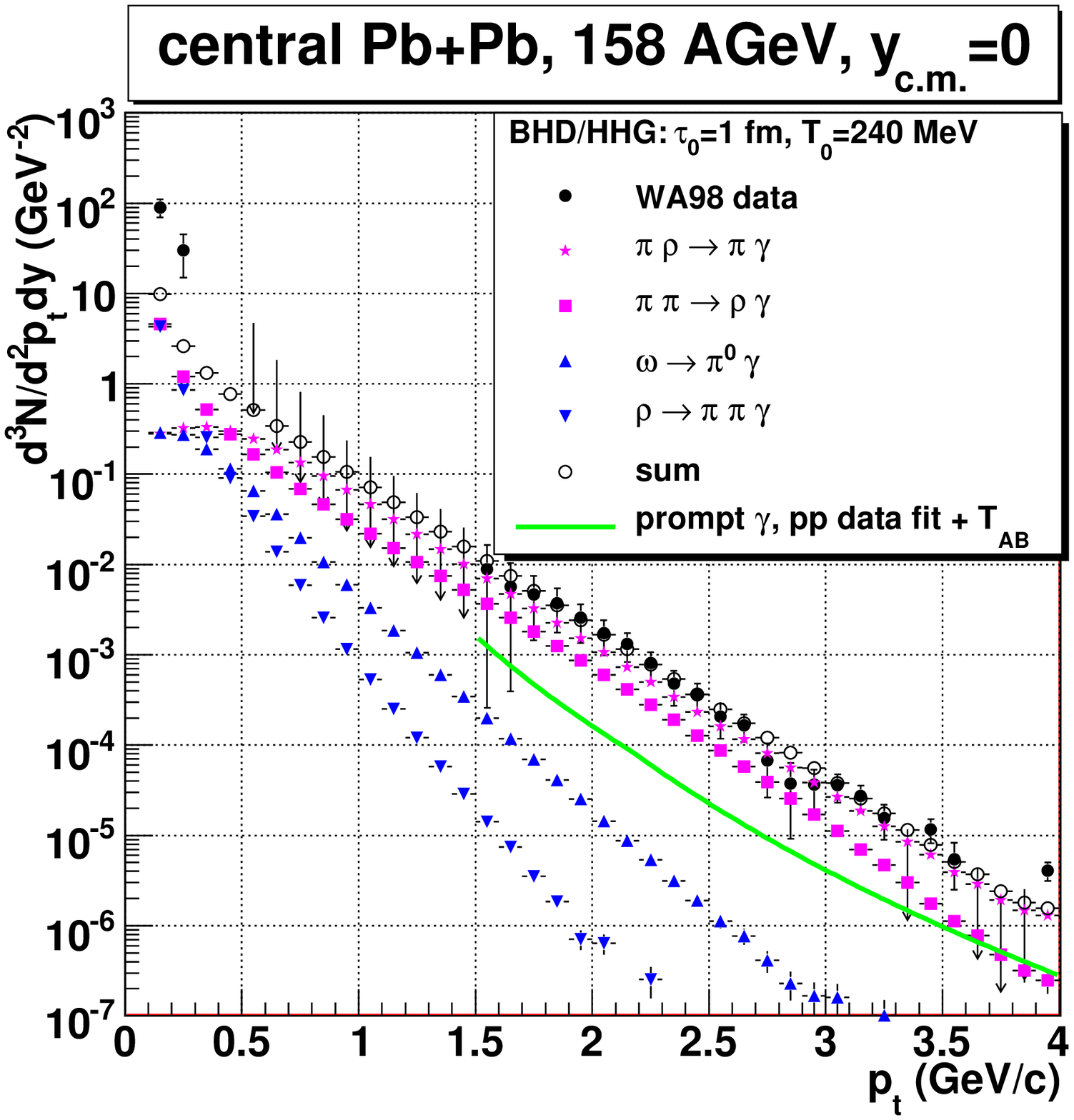}}
\caption{Direct photon spectrum for central Pb+Pb collisions at SPS}
\label{fig:GeTPHHG-SPS}
\end{figure}

\begin{figure}[p]
\centerline{
\includegraphics[width=100mm,height=100mm]{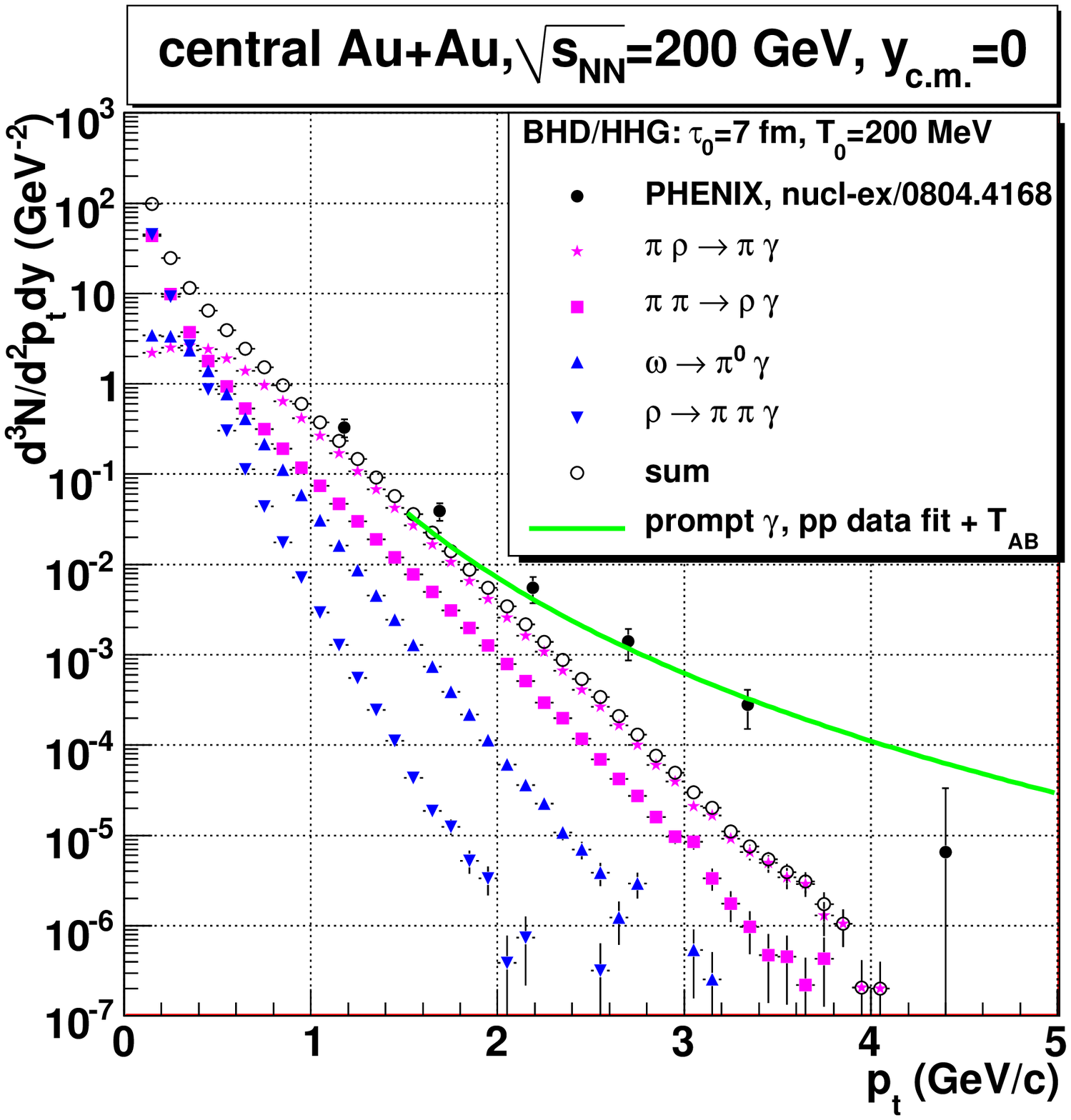}}
\caption{Direct photon spectrum for central Au+Au collisions at RHIC}
\label{fig:GeTPHHG-RHIC}
\end{figure}

Figs.~\ref{fig:GeTPHHG-SPS} and ~\ref{fig:GeTPHHG-RHIC} show comparison of SPS~\cite{WA98data} and RHIC~\cite{PHENIXdata} data with results of the generator GeTP-HHG.C realized as a macro. 
SPS data can be fitted by the BHD with the HHG scenario, except low $p_{t}$ points. For fitting the RHIC data unreasonable values of
parameters should be used, too large $\tau_{0}$=7 fm/c, and small $T_{0}$=200 MeV. The HHG scenario is hardly realized at the RHIC energy.
At the RHIC energy prompt photons dominate science $p_{t}$=2 GeV/c while at SPS they are not important up to 4 GeV/c.

\subsection{Generator for the QGP scenario}
In the QGP scenario of high-energy heavy-ion collisions three phases are assumed: a pure QGP phase, a mixed phase 
(QGP and HHG coexist) and a pure HHG phase. Ideal baryon free massless parton gas approximation for QGP
and ideal massless meson gas approximation for HHG are explored. 
A first order phase transition at the critical temperature $T_{c}$ is assumed. Other parameters of the QGP scenario are
the number of degree of freedom $g_{q}$ in QGP defined by the number of colors, $N_{c}$, and flavors, $N_{f}$, and effective number 
of degree of freedom $g_{h}$ in HHG. Time moments between the pure QGP and the mixed phase, $\tau_{c}^{q}$, between the mixed phase
and the pure HHG, $\tau_{c}^{h}$, and time of the freeze-out $\tau_{f}$ are expressed through the parameters.

The photon emission rates for QGP have been evaluated in perturbative thermal QCD applying hard thermal loop (HTL) resummation.
Contribution of the next to leading order diagrams (bremsstrahlung and annihilation with scattering) is the same order in $\alpha_{s}$
as from the leading order diagrams (Compton scattering and $q\bar q$ annihilation). This means that thermal photon production in
QGP is a non-perturbative mechanism that can not be accessed in perturbative HLT resummed thermal field theory. One must consider the
obtained QGP rates as an educated guess.

\begin{figure}[p]
\centerline{
\includegraphics[width=100mm,height=100mm]{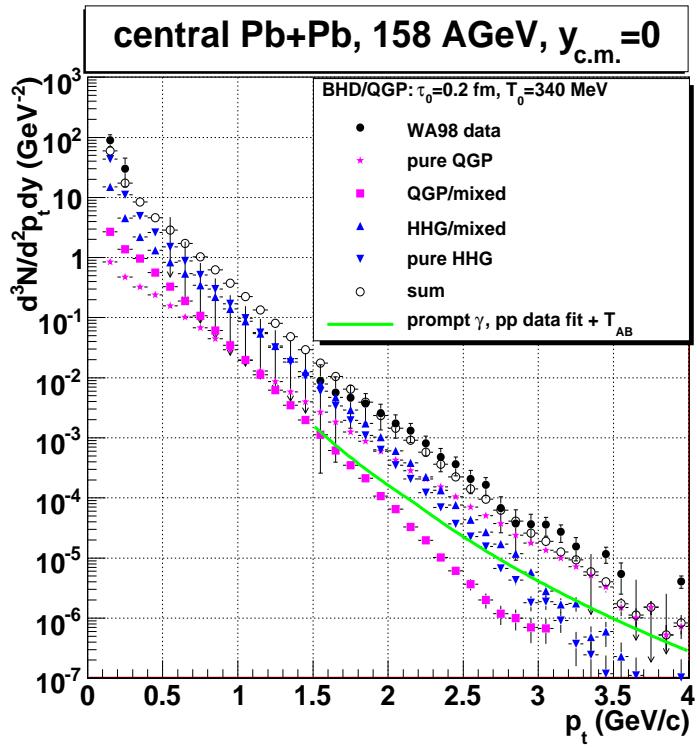}}
\caption{Direct photon spectrum for central Pb+Pb collisions at SPS}
\label{fig:GeTPQGP-SPS}
\end{figure}

\begin{figure}[p]
\centerline{
\includegraphics[width=100mm,height=100mm]{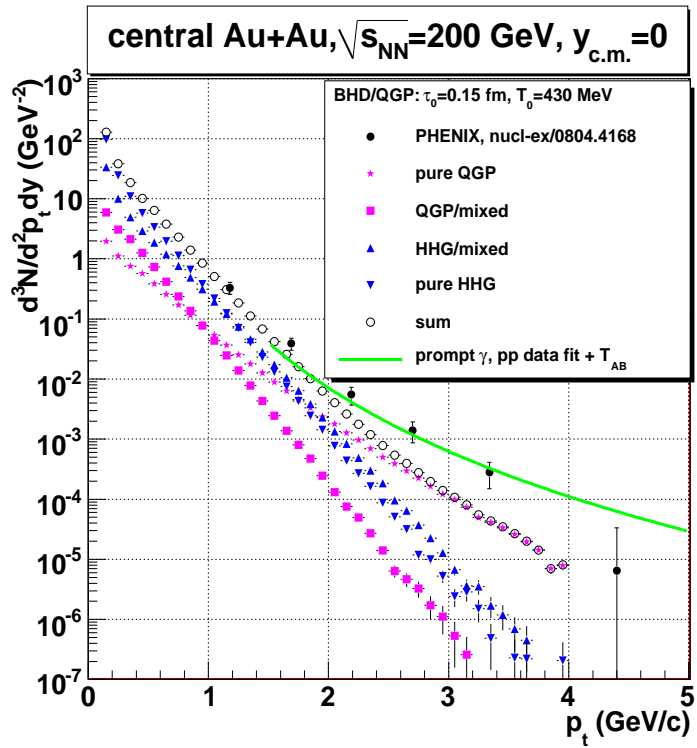}}
\caption{Direct photon spectrum for central Au+Au collisions at RHIC}
\label{fig:GeTPQGP-RHIC}
\end{figure}

Figs.~\ref{fig:GeTPQGP-SPS} and ~\ref{fig:GeTPQGP-RHIC} show comparison of the SPS and RHIC data with results of the generator GeTP-QGP.C 
realized as a macro (parameter values $N_{c}$=3, $N_{f}$=3, $T_{c}$=170 MeV and $g_{h}$=8 have been used). 
With the QGP scenario one can try to describe also low $p_{t}$ SPS data.
The RHIC data can be reproduced with reasonable values of $\tau_{0}$ and $T_{0}$.
QGP outshines HHG at $p_{t}>$2 GeV/c. One can compare the BHD spectra with 2+1 hydrodynamics results~\cite{2+1RHIC}.
While spectrum from QGP is almost the same the HHG spectrum is steeper. One if the reasons is the radial flow in the 2+1 case.

\begin{figure}[h]
\centerline{
\includegraphics[width=100mm,height=100mm]{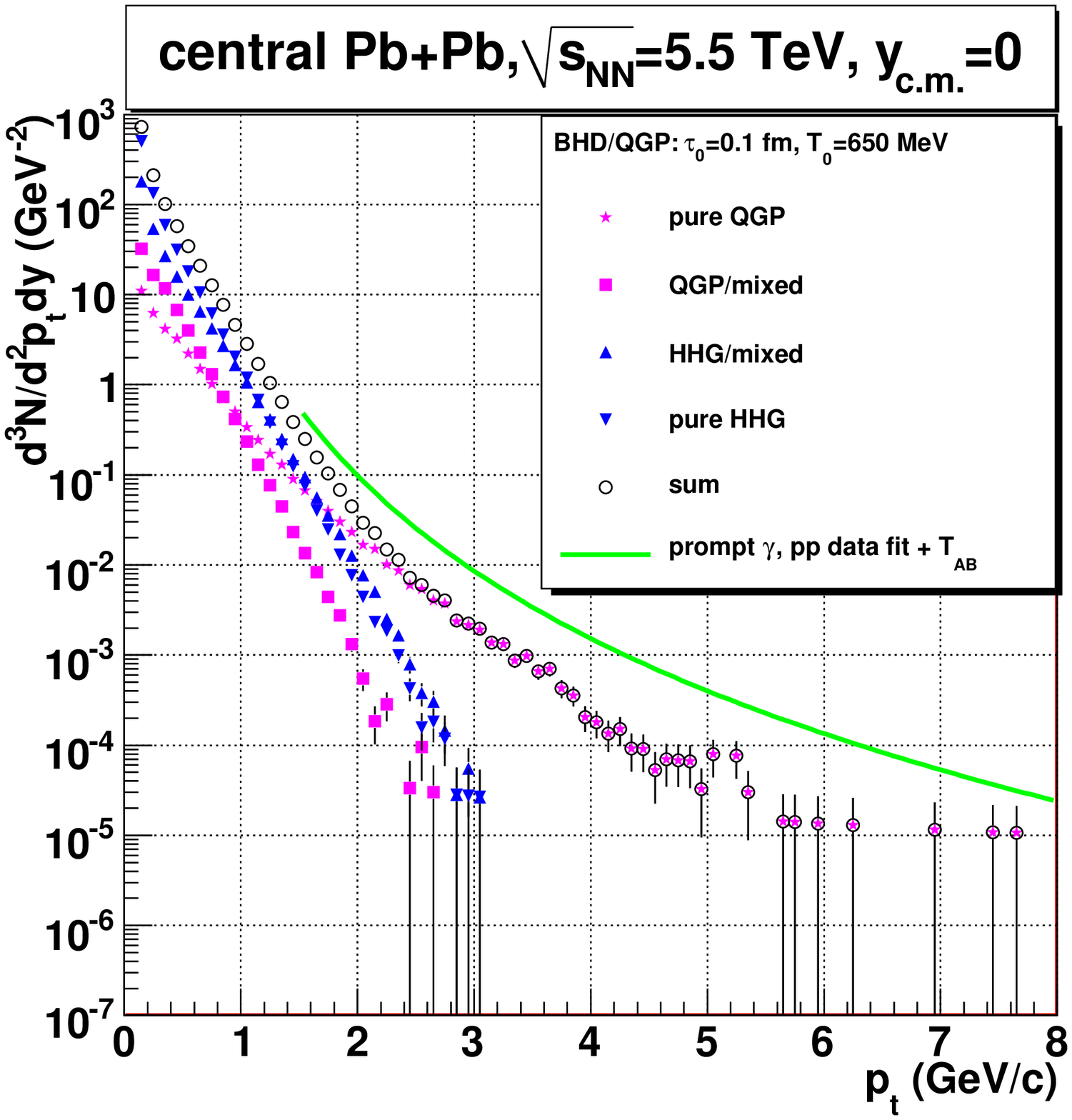}}
\caption{Direct photon spectrum for central Pb+Pb collisions at LHC}
\label{fig:GeTPQGP-LHC}
\end{figure}

Fig.~\ref{fig:GeTPQGP-LHC} demonstrates predictions for the LHC energy. Parameter values $\tau_{0}$=0.1 and $T_{0}$=650 MeV have been
chosen the same as in 2+1 hydrodynamics~\cite{2+1LHC}.
QGP outshines HHG at $p_{t}>$2 GeV/c while in the 2+1 hydrodynamics it happens at 3 GeV/c.

Table~\ref{tab:dNdy} summarizes results for SPS, RHIC and LHC energies.
\begin{table}[tbp]
\begin{center}
\begin{tabular}{|c|c|c|c|c|c|c|c|}
\hline
$\sqrt{s}$&$T_{0}$&$\tau_{0}$&$\tau_{c}^{q}$&$\tau_{c}^{h}$&$\tau_{f}$&$dN_{\gamma}/dy$& INIT \\
    GeV   &  MeV  &   fm/c   &    fm/c      &     fm/c     &   fm/c   &                & CPU  \\
\hline
     17   &   340 &   0.20   &     1.6      &      9.5     &    46.7  &       14       & 110 s\\
\hline
    200   &   430 &   0.15   &     2.4      &     14.4     &    70.8  &       31       & 160 s\\ 
\hline
   5500   &   650 &   0.10   &     5.6      &     33.2     &    163   &      173       & 390 s\\
\hline
\end{tabular}
\end{center}
\caption{Characteristics of central collisions at SPS, RHIC and LHC energies in the QGP scenario}
\label{tab:dNdy}
\end{table}
Except the initial and time evolution parameters the rapidity densities of thermal direct photons, $dN_{\gamma}/dy$,
are pointed out. Though QGP dominates at high $p_{t}$ its contribution in $dN_{\gamma}/dy$ is ~10\%.
In the last column of the table typical CPU time needed to calculate $dN/dp_{t}$ photon distribution used to
randomly chose a $p_{t}$ value.

\section{Summary}
Three fast generators of direct photons in the central rapidity region of high-energy heavy-ion collisions have been presented 
The generator of prompt photons is based on a tabulation of $p+p(\bar p)$ data and binary scaling.
Two generators of thermal direct photons, for HHG and QGP scenarios,
assume the 1+1 Bjorken hydrodynamics. SPS and RHIC data can be fitted better by scenario with QGP. Predictions for the 
LHC energy have been made. The generators have been realized as macros for the ROOT analysis package.
First two of them have been implemented into the FASTMC code of the UHKM package~\cite{UHKM}.

\section{Acknowledgments}
This  work  was supported by the Russian Foundation for Basic Research 
(grant numbers 06-08-01555, 08-02-00676, 08-02-92496), INTAS (grant number 06-1000012-8914) and Federal agency of Russia for atomic energy (Rosatom).

\end{document}